\documentclass[pra, aps, showpacs, reprint]{revtex4-2}

\usepackage[utf8]{inputenc}
\usepackage{graphicx}
\usepackage{amsmath}
\usepackage{amssymb}
\usepackage{braket}
\usepackage{color}
\usepackage{siunitx}
\usepackage[caption=false]{subfig}
\usepackage{amsmath}
\usepackage{verbatim}
\usepackage{tikz}
\usetikzlibrary{positioning}
\usepackage[draft]{hyperref}
\usepackage[noabbrev,capitalise]{cleveref}

\newlength{\colfigwidth}
\newlength{\figh}
\newlength{\figl}
\renewcommand\vec[1]{\ensuremath\boldsymbol{#1}}
\newcommand{\uvec}[1]{\hat{\mathbf{#1}}}

\newcommand{\diff}[1]{\mathrm{d} #1}

\newcommand{\partialddx}[1]{\frac{\partial}{\partial #1}}

\definecolor{purpleplot}{RGB}{137, 48, 141}

\begin{document}

\title{Spin-transfer-assisted parametric pumping of magnons in yttrium iron garnet 
}

\author{Therese Frostad}
\author{Hans L. Skarsv\aa g}
\author{Alireza Qaiumzadeh}
\author{Arne Brataas}
\affiliation{
Center for Quantum Spintronics, Department of Physics, NTNU Norwegian University of Science and Technology, NO-7491 Trondheim, Norway
}

\date{\today}

\begin{abstract}
The combination of parametric pumping and spin-transfer torque is a powerful approach that enables high-level control over magnetic excitations in thin-film ferromagnets. 
The excitation parameters, such as pumping power and external field strength, affect the instabilities of individual magnon modes. 
We theoretically explore how the simultaneous effects of parametric pumping and spin transfer torque influence these magnetic instabilities in a thin-film ferromagnet. Within the Landau-Lifshitz-Gilbert framework, we perform micromagnetic simulations of magnon excitations in yttrium iron garnet by pumping, spin transfer torque, and a combination of the two. 
We find that consistent with experimental results, the magnitude and direction of the spin-transfer torque tune the parametric instability thresholds.

\end{abstract}

\maketitle

\section{Introduction}

Spin waves function as information carriers in spintronics. Magnons, the quanta of spin waves, may be created in ferromagnetic materials through numerous methods. Here, we address two common approaches for exciting the magnetization dynamics of thin-film ferromagnets.

\textit{Parametric pumping} is a conventional nonlinear excitation method that uses a microwave field \cite{morgenthaler1960survey, schlomann1960recent}. There is an applied AC magnetic field in the same direction as the external static magnetic field orienting the magnetization during parallel parametric pumping. The alternating pumping field induces oscillations in the ferromagnetic spin system. When the strength of the pumping field exceeds a certain threshold, an instability occurs that creates two magnons with opposite momenta. Energy conservation ensures that the two magnons have a lower frequency than the driving microwave field. Parametric pumping was introduced in the 1950s \cite{anderson1955instability,suhl1956nonlinear,suhl1957theory}. In 2006, parametric pumping generated Bose-Einstein condensation (BEC) of magnons in yttrium iron garnet (YIG) at room temperature \cite{demokritov2006bose}. This work motivated further discussions on magnon creation by pumping, thermalization processes of the created magnons, and the coherence of the emerging condensate \cite{demokritov2008quantum,demidov2008observation,tupitsyn2008stability,rezende2009theoryCoherence,rezende2009theorySuperradiance,nowik2012spatially,serga2014bose,clausen2015stimulated,dzyapko2016high,hahn2020collisionless,mohseni2020parametric,hayashi2018spin,heinz2021parametric,noack2019enhancement, hahn2021effect}. 
Parametric pumping was further investigated as a tool for spin wave amplification \cite{verba2018amplification, bracher2014parallel}. 

\textit{Spin-transfer torque (STT)} is a more recent approach to create magnons \cite{ralph2008spin,brataas2012current}. An external current or voltage induces a magnetization torque. One possible realization is to pass an electrical current through an adjacent metal such that the spin-Hall effect creates a spin accumulation in the metal. The spin accumulation may subsequently generate a torque on the magnetization in the neighboring ferromagnet. The STT may be employed to either inhibit or assist magnon creation because the torque has a component that acts damping-like or anti-damping-like, depending on the sign of the spin accumulation \cite{urazhdin2010parametric, edwards2012parametric, hamadeh2014full, tserkovnyak2002enhanced,wang2011control}. This control over the damping can create spin-torque oscillators. In this case, the current or voltage controls the oscillator frequency. STT can also be used to switch the magnetization configuration \cite{slonczewski1996current,myers1999current,katine2000current}. The latter feature enables concepts for magnetic random access memories. Recently, STT was utilized to create BEC in thin-film BiYIG\cite{divinskiy2021evidence}. 

Combining parametric pumping and STT achieves high-level control over magnon creation. It is known that STT tunes the instability thresholds of the parametric pumping mechanism \cite{urazhdin2010parametric, edwards2012parametric, lauer2016spin}. Lauer {\it{et al}}. \cite{lauer2016spin} merged parallel parametric pumping and STT on a thin ($\SI{100}{\nano\metre}$) YIG film. They monitored the resulting magnetization dynamics by Brillouin light scattering spectroscopy. The experiment revealed that the STT tunes the effective damping, thereby changing the threshold pumping power required to excite magnons by parametric pumping. Motivated by this experimental study, we conduct a large-scale micromagnetic simulation of a similar system. We apply both a parallel pumping field and STT. We determine the temporal evolution of the magnetization, from which we determine the stability phase diagram of spin-transfer-assisted parametric pumping. We also resolve the wavevector dependence on the nonequilibrium population of magnons.

We organize the remainder of this paper as follows. First, \cref{sec:magdyn} introduces the  theoretical framework. Therein, we present the Landau-Lifshitz-Gilbert (LLG) equation that describes the magnetization dynamics of the thin YIG film. \cref{sec:num} presents the setup for our micromagnetic simulations. To separate the effects of magnon excitation by parametric pumping and STT, we perform different simulations for the two phenomena. Finally, we simulate the combination of parametric pumping and STT applied to the YIG film. \cref{sec:numres} presents the results, and \cref{sec:conc} summarizes our findings.

\section{\label{sec:magdyn}Magnetization dynamics}

We calculate the time evolution of the unit vector along the magnetization direction $\vec{m}(\vec{r},t)$ within the LLG framework. The LLG equation reads 
\begin{equation}\label{eq:llg}
 \dot{\vec{m}} = 
- \gamma (\vec{m} \times \vec{H}_\text{eff}) 
+ \alpha(\vec{m} \times \dot{\vec{m}}) \, . 
\end{equation}
Here, $\gamma = \SI{1.7595e11}{\radian\per\tesla\per\second}$ is the gyromagnetic ratio, and $\alpha$ is the dimensionless Gilbert damping constant. The effective field $\vec{H}_\text{eff}$ includes contributions from the exchange field $\vec{H}_\text{exch}$, the external magnetic field $\vec{H}_\text{ext}$, and the dipole-dipole field $\vec{H}_\text{d-d}$. In our free energy we disregard the crystalline anisotropy since it is very small in YIG. 

The exchange field arises from the Heisenberg exchange interaction between neighboring spins, and it promotes a homogeneous magnetization,  
\begin{equation}
\vec{H}_\text{exch} 
= 2 \frac{A_\text{ex}}{M_S} \nabla^2 \vec{m} \, .
\end{equation}
Here, $A_\text{ex}$ is the exchange stiffness, and $M_S$ is the saturation magnetization.

We can orient the magnetization in YIG along a desired direction by applying a sufficiently high static field of strength $H_0$. We define a film in the ($x$,$y$)-plane, and choose to apply the static field in the $\uvec{x}$-direction, 
\begin{equation}
\vec{H}_\text{ext} = H_{0}\uvec{x} \, . 
\end{equation}
The external field term may include additional dynamic fields applied to the film. 

The dipole-dipole field consists of the static demagnetization field and dynamic terms due to the long-range magnon-magnon interactions. In general, the dipole field at position $\vec{r}$ can be expressed in terms of contributions from the magnetization $\vec{m}(\vec{r'})$ at distance $\vec{d}=\vec{r}-\vec{r'}$ integrated over the film volume $V$,
\begin{equation}
\vec{H}_\text{d-d}(\vec{r}) 
= \frac{\mu_0}{4 \pi} M_S\int_{V}^{} \diff{\vec{r'}}
\frac{3 (\vec{m}(\vec{r'}) \cdot \vec{d})\vec{d}}{|\vec{d}|^5} 
- \frac{\vec{m}(\vec{r'})}{|\vec{d}|^3}
\, . 
\end{equation}
Here, $\mu_0$ is the permeability of free space.
The dipole field can be expressed in terms of a $3 \times 3$ tensor $\hat{G}(\vec{r},\vec{r'})$ \cite{kalinikos1986theory, kalinikos1981spectrum}
,
\begin{equation}\label{eq:diptensor}
\vec{H}_\text{d-d}(\vec{r}) = \mu_0 M_S\int_{V}^{} \diff{\vec{r'}}
\hat{G}(\vec{r},\vec{r'}) \cdot \vec{m}(\vec{r'})
\, .
\end{equation}
The tensor elements of $\hat{G}(\vec{r},\vec{r'})$ are $G_{\alpha \beta} = - \frac{1}{4 \pi} \partialddx{\alpha} \partialddx{\beta} \frac{1}{|\vec{d}|}$, for $\alpha, \beta = x,y,z$.

\begin{figure}[ht]
\begin{tikzpicture}
\node (img)  {
\includegraphics[width=0.9\colfigwidth]{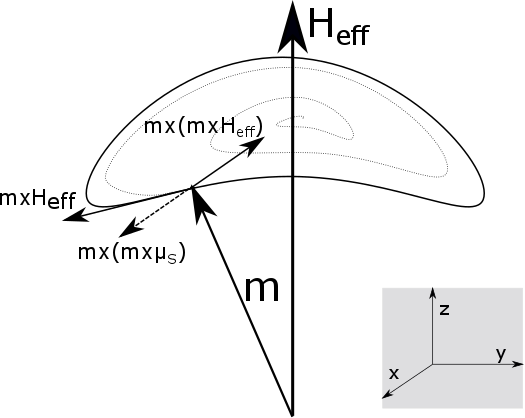}
};
\node[below=of img, node distance=0cm, font= \color{black}, fill=gray!25, rounded corners=4pt, xshift = 2.6cm, yshift=2.75cm] {$\mathbf{x}$};
\node[below=of img, node distance=0cm, font= \color{black}, fill=gray!25, rounded corners=4pt, xshift=1.87cm, yshift=2.065cm] {$\mathbf{y}$};
\node[below=of img, node distance=0cm, font= \color{black}, fill=gray!25, rounded corners=4pt, xshift=3.25cm, yshift=2.237cm] {$\mathbf{z}$};
\end{tikzpicture}
\caption{\label{fig:precession} 
The magnetization $\vec{m}$ precesses around $\vec{H}_\text{eff}$ due to the field-like torque $-\vec{m} \times \vec{H}_\text{eff}$. The damping $-\vec{m} \times (\vec{m} \times \vec{H}_\text{eff})$ can be tuned by applying an STT. The torque acts as a damping-like or anti-damping-like torque depending on the sign of the spin accumulation $\mu_S$. The precession in the ($y$,$z$)-plane is elliptical due to the dipole contributions from the finite film thickness in the $\uvec{z}$-direction.
}
\end{figure}
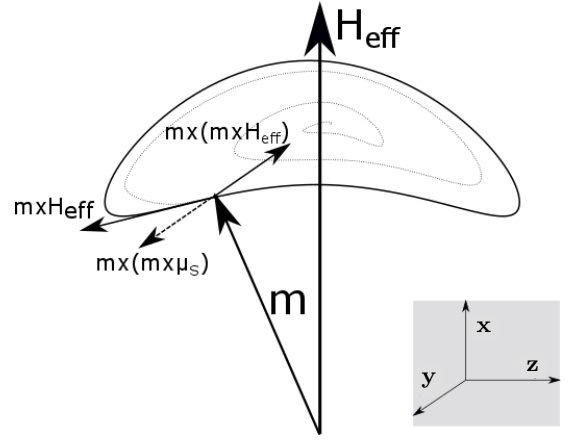

\subsection{Spin-wave spectra}

We now employ the thin-film approximation by assuming that the magnetization is uniform in the $\uvec{z}$-direction.  Therefore, the thin film is effectively two dimensional. In the expression for the dipole field \eqref{eq:diptensor}, we average across the variation in the $\uvec{z}$-direction by integrating over the film thickness $L_z$. The total effective field of the LLG equation becomes a surface integral over the in-plane coordinate $\vec{\rho} = x\uvec{x} + y\uvec{y}$,
\begin{align}
\vec{H}_\text{eff}(\vec{\rho}, t) =
H_0\uvec{x} 
+ 2 \frac{A_\text{ex}}{M_S} \nabla_{\vec{\rho}}^2 \vec{m}(\vec{\rho}, t) \nonumber \\ 
+ \mu_0 M_S \int_{S} \hat{G}(\vec{\rho},\vec{\rho'}) \vec{m}(\vec{\rho'}, t) \diff{\vec{\rho'}} \, ,
\label{eq:heff}
\end{align}
where the effective two-dimensional dipole tensor becomes
\begin{equation}
\hat{G}(\vec{\rho},\vec{\rho'}) = \frac{1}{L_z} \int_{-L_z/2}^{L_z/2} \diff{z} \int_{-L_z/2}^{L_z/2} \diff{z'}
\hat{G}(\vec{r},\vec{r'})
\, .
\end{equation} 
We will consider the nonlinear response of the magnetization to the parallel pumping field in our numerical investigations presented below. However,  relating our results to the linear response regime is also instructive. 

In the linear response regime, the magnetization direction is $\vec{m}(\vec{\rho}, t) = \uvec{x} + \delta \vec{m}(\vec{\rho}, t)$, where the small out-of-equilibrium deviation $\delta \vec{m}(\vec{\rho}, t)$ lies in the ($y$,$z$)-plane, as shown in \cref{fig:precession}. Additionally, we assume the magnetization is precessing with frequency $\omega$; thus, $\delta \vec{m}(\vec{\rho}, t) = \delta \vec{m}(\vec{\rho}) e^{i\omega t}$. We insert the effective field into the LLG \cref{eq:llg}. In the linear response regime, we retain only the first-order terms in the deviation $\delta \vec{m}(\vec{\rho}, t)$. We proceed to define the spatial Fourier transforms,
\begin{equation}
\vec{m}(\vec{k}) = \mathcal{F}[\vec{m}(\vec{\rho})] =
\frac{1}{2\pi} \int \vec{m}(\vec{\rho}) e^{-i \vec{k}\cdot\vec{\rho}} \diff{\vec{\rho}}
\label{eq:fourierdef1}
\, ,
\end{equation}
\begin{equation}
\vec{m}(\vec{\rho}) = 
\int \vec{m}(\vec{k}) e^{-i \vec{k}\cdot\vec{\rho}} \diff{\vec{k}}
\, . 
\end{equation}
Here, $\vec{k} = |k| (\cos \theta_{k} \uvec{x} + \sin \theta_{k} \uvec{y})$ is the magnon wavevector. We can express the LLG equation as an eigenvalue problem, where the precession frequencies are the eigenvalues. We may then obtain the well-known dispersion relation for an extended thin film \cite{kalinikos1986theory,kreisel2009microscopic,ruckriegel2015rayleigh},
\begin{align}
\omega(k, \theta_{k}) 
&=  \sqrt{\omega_H + \omega_M l_\text{ex}^2 k^2 + \omega_M(1-f_{k})\sin^2 \theta_{k}} \nonumber \\
& \times \sqrt{\omega_H + \omega_M l_\text{ex}^2k^2 + \omega_M f_{k}} \, .
\label{eq:dispersionrel}
\end{align}
Here, we have defined $\omega_M= \gamma \mu_0 M_S $, $\omega_H= \gamma \mu_0 H_0$ and the magnetic exchange length $l_\text{ex} = \sqrt{2A_\text{ex}/\mu_0 M_S^2}$. The form factor $f_{k}$ accounts for the film thickness,
\begin{equation}
f_{k} 
= \frac{1 - e^{-|k| L_z}}{|k| L_z} 
\, . 
\end{equation}

\begin{figure}[ht]
\centering
\includegraphics[width=\colfigwidth]{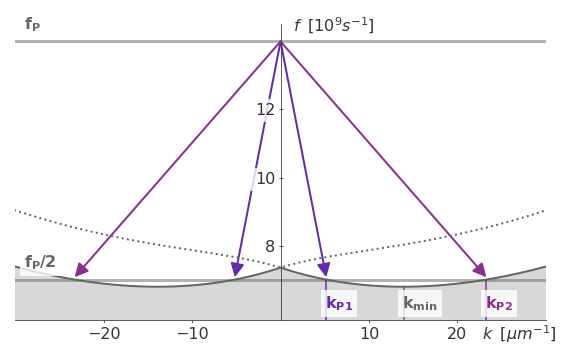}
\caption{\label{fig:dispplot} 
The dispersion relation for spin waves traveling in the $\uvec{x}$-direction of a thin YIG film ($k_x$-branch), as given by \cref{eq:dispersionrel}. Spin waves traveling in the $\uvec{y}$-direction have higher energies, as indicated by the dashed line ($k_y$-branch). During parametric pumping, two magnons are created at wavevectors of opposite signs. 
These magnons both have frequency $\omega_p/2$. 
In the $k_x$-branch, these magnons are found at $\pm k_{p1}$ and $\pm k_{p2}$. In turn, they thermalize and eventually occupy the dispersion minimum at $k_\text{min}$. The parameters are from \cref{tab:yigparamstable,tab:simparamstable}, with $\mu_0 H_0 = \SI{190}{\milli \tesla}$. 
}
\end{figure}

The spin waves traveling in the $\uvec{x}$-direction ($\theta_{k} = 0$) have the lowest energy. This lower magnon branch is shown in \cref{fig:dispplot}. It is symmetric in $k$, and has a double minimum at $k=\pm k_\text{min}$. The exchange energy controls the spin waves at high wavevectors $k\gg l_\text{ex}^{-1}$ so that the dispersion is quadratic in $k$. On the other hand, the dipole interaction controls the spin waves at small wavevectors $k< L_z^{-1}$. 

\subsection{Magnon excitation}

We proceed to examine magnon excitation by parametric pumping. In the parallel pumping geometry, the magnetization is in-plane and oriented along the $\uvec{x}$-axis by a static field of magnitude $H_0$. There is an additional alternating microwave field $\vec{h}_{p}(t)$ parallel to the static field that functions as a magnon pump. The external field has static and oscillating contributions, 
\begin{equation}
\vec{H}_\text{ext} = (H_{0} + h_{p}\sin(\omega_p t))\uvec{x}
\, . 
\end{equation}
Here, $h_{p}$ is the amplitude and $\omega_p$ is the frequency of the oscillating pumping field. When the strength of the pumping field exceeds a critical threshold, $h_p > h_p^\text{crit}(H_0)$, the magnetization starts to precess around the $\uvec{x}$-axis. For a thin film, this precession is elliptical due to the dipole interactions from the finite thickness in the $\uvec{z}$-direction. By energy conservation, the created magnons have half the frequency of the pumping field, $\omega_{\vec{k}} = \omega_{\vec{-k}} = \omega_p/2$, and are opposite wave vectors, as illustrated in \cref{fig:dispplot}. 
This is the oscillation of the x-component of the magnetic moments that couple to the magnetic pulse in the parallel pumping mechanism. Since dipolar-magnons in YIG are elliptically polarized, the x component of the magnons oscillates, and therefore parallel pumping is possible. Parametric pumping mainly excites elliptical dipole-dominated low energy magnon modes. We therefore expect that magnons are pumped in the $k_x$-branch, at points $k_x=\pm k_{p1}$, as illustrated in \cref{fig:dispplot}. Following this reasoning, the most efficient pumping occurs when $k_{p1}=k_{p2}=0$ at ferromagnetic resonance (FMR) conditions, since the ellipticity is highest in that situation. For a fixed pumping frequency $\omega_p$, the external magnetic field strength $H_\text{FMR}$ corresponding to the resonance frequency $\omega_\text{FMR}$ is approximated by the \textit{Kittel formula} \cite{kittel1948theory}, as expected from \cref{eq:dispersionrel} in the limit $k \rightarrow 0$, 
\begin{equation} 
\omega_\text{FMR} = \omega_p/2 = \gamma\sqrt{H_\text{FMR}(H_\text{FMR} + \mu_0 M_S)} \, .
\label{eq:kittel}
\end{equation} 

Next, we consider how the STT affects the magnetization dynamics. In thick films, the effective volume of surface modes is smaller than that of volume modes. Therefore, exciting surface modes by STT is easier than exciting bulk modes by STT \cite{xiao2012spin,Kapelrud:prl2013}. However, we consider a thin film without any surface anisotropy, where the magnetization is uniform along the thickness direction. Surface modes are thus less important in this case, and we will not discuss them further.

\begin{figure}[ht]
\centering
\includegraphics[width=\colfigwidth]{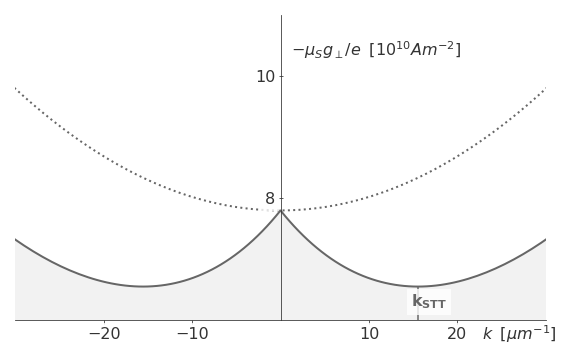}
\caption{\label{fig:spinaccplot} 
The critical spin accumulation $-\mu_S^\text{crit}g_{\perp}/e$ for exciting spin waves traveling in the $\uvec{x}$-direction of a thin YIG film ($k_x$-branch), as given by \cref{eq:mucrit}. Spin waves traveling in the $\uvec{y}$-direction require a higher spin accumulation, as indicated by the dashed line ($k_y$-branch).
The parameters are from \cref{tab:yigparamstable,tab:simparamstable}, with $\mu_0 H_0 = \SI{100}{\milli \tesla}$. For these parameters, we find that the first magnons are excited at $\theta_k=0$, $|\vec{k}_\text{STT}| \approx \SI{15.5}{\per \micro \meter}$, for the critical spin accumulation $\mu_S^\text{crit}(\vec{k}_\text{STT}) g_{\perp}/e \approx -\SI{6.55e10}{\ampere \per \meter \squared}$. 
}
\end{figure}

The STT typically arises from a spin accumulation in an adjacent normal metal. We assume that a charge current in the adjacent metal layer (e.g., Pt) produces the spin accumulation polarized along the $\uvec{x}$-axis. The resulting torque on the YIG interface is expressed in terms of the spin-mixing conductance per area $g_{\perp}~[1/\Omega\text{m}^2]$  and spin accumulation density $\vec{\mu}_S = \mu_S \uvec{x}$,
\begin{align}
 \dot{\vec{m}} & = 
- \gamma (\vec{m} \times \vec{H}_\text{eff})
+ \alpha(\vec{m} \times \dot{\vec{m}}) \nonumber \\
& - \frac{\gamma \hbar}{2 e^2 L_z M_S} g_{\perp} \vec{m} \times (\vec{m} \times \vec{\mu_S}) \, . 
\label{eq:LLGSTT}
\end{align}
Here, $e$ is the elementary charge, and $\hbar$ is the reduced Planck constant. In our geometry, a positive (negative) spin accumulation results in a damping-like (anti-damping-like) STT. We proceed to find an expression for the critical spin accumulation $\mu_S^\text{crit}$ at which the damping is overpowered. As before, the effective field in \cref{eq:heff} includes the static field, exchange interaction and dipole interactions. Inserting the effective field into \cref{eq:LLGSTT}, the imaginary part of the eigenfrequencies determines the critical spin accumulation required to excite magnons with a specific wavevector, 
\begin{align}
\mu_S^\text{crit}(k, \theta_{k}) 
&= -\frac{2 L_z e^2 \alpha M_S}{\hbar g_{\perp} \gamma}
\Big( \omega_H + \omega_M l_\text{ex}^2 k^2 \nonumber \\
&+ \frac{1}{2}\omega_M (f_{k}+ (1-f_{k}) \sin^2 \theta_k ) \Big) \, .
\label{eq:mucrit}
\end{align}
\cref{eq:mucrit} can be minimized to yield the first magnon wavevector to be excited by STT. This wave vector denoted by $\vec{k}_\text{STT}$ is in the $k_x$-branch ($\theta_k=0$), as illustrated in \cref{fig:spinaccplot}.
We note that the threshold spin accumulation in \cref{eq:mucrit} is linear in the external field strength.

\section{\label{sec:num}Numerical approach}

We consider the magnetization dynamics in a thin film of YIG. First, we define a square film with lateral side lengths $L_{x,y}$ in the ($x$,$y$)-plane. The film thickness $L_z=\SI{100}{\nano\meter}$ is much smaller than the lateral side lengths. We use the thin-film approximation in which there is no magnetization variation in the $\uvec{z}$-direction. Laterally, we divide the film into $N= n_{x,y}^2 $ cells, each of size $l_{x,y}^2 \times L_z$. The unit vector in the magnetization direction of each cell is $\vec{m}(\vec{\rho}_i, t)$. We find the time evolution of each magnetization vector by solving the LLG equation at successive time intervals. 

The simulations provide full information on the local magnetization within each cell of the film. This information allows the definition of a relative magnon density $\eta$ in terms of the longitudinal component of the average magnetization, 
\begin{equation}
\eta = 1 - \langle \vec{m}_x \rangle
\label{eq:magdensity}
\, ,
\end{equation} 
where $\langle \vec{m}_x \rangle = \sum_{i=1}^{i=N} \vec{m}(\vec{\rho}_i,t)/N$. We can find the magnon distribution $\zeta(k_x,k_y)$ as a function of the wavevector evaluating the Fourier transform of the transverse components of the magnetization, 
\begin{equation}
\zeta(k_x, k_y) = |\mathcal{F}[m_y(\vec{\rho}, t)]|^2 + |\mathcal{F}[m_z(\vec{\rho}, t)]|^2
\label{eq:magdistribution}
\, .
\end{equation} 
Here, $\mathcal{F}[~]$ denotes a discrete Fourier transform defined similar to the continuous transform in \cref{eq:fourierdef1}. 

Our simulations include magnon excitation by parallel pumping, the STT, and the combination of the two. To this end, we use the open-source GPU-accelerated software \textit{MUMAX3}.\cite{vansteenkiste2014design} The simulations start from an initial magnetization state $\vec{m}(\vec{\rho}_i, t=0)$, which must deviate from the uniform state for the magnetization dynamics to start. We apply the pumping fields and/or the STT over the entire surface of the film, and we investigate the resulting magnon density as a function of the bias field $H_0$, the pumping field amplitude $h_p$ and the spin accumulation $\mu_S$. \cref{sec:numres} provides more details on the initialization and parameter variation of each simulation. Although the lateral size of the film is much larger than the film thickness, we expect some finite size effects. The material parameters for YIG are listed in \cref{tab:yigparamstable}. The damping coefficient $\alpha = 10^{-2}$ is set higher than, e.g., Ref.~\cite{lauer2016spin} ($\alpha \sim 10^{-4}$), such that the magnetization may reach steady state at acceptable running times. We detail the typical numerical parameters of the simulations in \cref{tab:simparamstable}. 

\begin{table}[h]
\caption{\label{tab:yigparamstable}Material parameters of YIG.}
\begin{ruledtabular}
\begin{tabular}{llr}
Parameter &Symbol &Value in SI \\
\hline
Saturation magnetization~\cite{lauer2016spin} 		& $\mu_0 M_S$ 	& $\SI{173}{\milli\tesla}$ 					\\
Exchange stiffness~\cite{klingler2014measurements}				& $A_\text{ex}$ & $\SI{3.65e-12}{\joule\per\metre }$ 		\\
Gilbert damping parameter		& $\alpha$ 		& $0.01$			
\end{tabular}
\end{ruledtabular}
\end{table}

\begin{table}[h]
\caption{\label{tab:simparamstable} System parameters.}
\begin{ruledtabular}
\begin{tabular}{llr}
Parameter &Symbol &Value in SI \\
\hline
YIG film thickness			& $L_z$     & $\SI{100}{\nano\metre}$   \\
YIG film lateral dimensions	& $L_{x,y}$ & $\SI{5}{\micro\metre}$    \\
Lateral number of cells 	& $n_{x,y}$ & $2^{9 }$                  \\ 
Lateral cell size			& $l_{x,y}$ & $\sim \SI{10}{\nano\metre}$ \\
Microwave pumping frequency	& $\omega_\text{p} = 2\pi f_p$ & $2\pi \times \SI{14}{\giga\hertz}$ \\
\end{tabular}
\end{ruledtabular}
\end{table}

\section{\label{sec:numres}Numerical results}

We discuss the numerical results of three scenarios: {\it{(i)}}) excitations performed exclusively by parallel pumping, {\it{(ii)}}) excitations performed exclusively by STT, and {\it{(iii)}}) excitations resulting from the combination of pumping and STT.

\subsection{\label{sec:numres_pp}Magnon excitation by parallel parametric pumping}

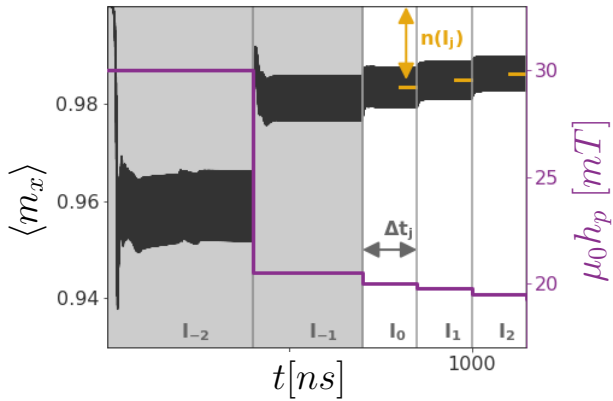
\begin{figure}[!htp]
\begin{tikzpicture}
\node (img)  {
\includegraphics[width=1.0\colfigwidth]{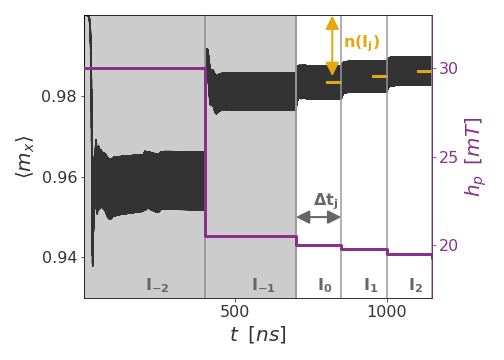}
};
\node[below=of img, node distance=0cm, font=\Large \color{black}, fill=white, 
yshift=2.05cm] {$t [ns]$};
\node[left=of img, node distance=0cm, rotate=90, anchor=center, font=\Large \color{black}, fill=white, 
xshift=2, yshift=-1.4cm] {$\langle m_x \rangle$};
\node[left=of img, node distance=0cm, rotate=90, anchor=center, font=\Large \color{purpleplot}, fill=white, 
xshift=2, yshift=-8.9cm] {$\mu_0 h_p \; [mT]$};
\end{tikzpicture}
\caption{\label{fig:mxplotmu5j0}
The temporal evolution of the spatial average of the magnetization $\langle m_x \rangle(t)$. At $t=0$, the magnetization randomly deviates from the uniform state. The shaded areas show the initialization intervals $I_{-2}$ and $I_{-1}$. During these intervals, the pumping is strong and creates an initial state. Thereafter, the pumping power $h_p$ decreases in intervals $I_j$ (purple). Within each interval, we compute the temporal average of the magnetization (orange) resulting in the magnon density $\eta(H_0, h_p)$. The presented data are an excerpt from the simulations at $\mu_0 H_0 = \SI{190}{\milli\tesla}$. 
}
\end{figure}

We first present simulations of magnon excitation by parallel parametric pumping. The number of created magnons is a measure of the efficiency of the pumping process. The simulations determine the relative magnon density as a function of the external field strength $H_0$ and pumping power $h_p$. In doing so, the simulations provide the critical threshold strength $h_p^\text{crit}(H_0)$ required to excite magnons. The threshold excitation strength may show hysteretic behaviour depending on whether we increase or decrease $H_0$ \cite{chen2017parametric, guo2014parametric}. For each fixed value of $H_0$, we run simulations in which $h_p$ is decreased in steps for each time interval of length $\Delta t$, as illustrated in \cref{fig:mxplotmu5j0}. 

Each simulation starts from a chosen initial magnetization state. A noisy input state is created by randomly pulling each magnetization vector slightly away from its uniform state. This contributes to the initial magnon density, as observed at $t=0$ in \cref{fig:mxplotmu5j0}. We proceed to strongly pump the system for two intervals $I_{-2}$ and $I_{-1}$. The gray shading in \cref{fig:mxplotmu5j0} shows these initialization intervals, which are not included  in the extracted results in \cref{fig:butterfly_weakStt}. After the initialization, we pump the systems in intervals $I_j$ ($j=0,1,2,...$) while decreasing the pumping strength for each interval. 

\cref{fig:mxplotmu5j0} shows how the magnetization reaches a steady state when the pumping is decreased. In the steady state, the magnetization precession is elliptical. The spatial average of the magnetization $\langle \vec{m}_x\rangle (t)$ determines the relative density of the magnons, as in \cref{eq:magdensity}.  We compute a time-averaged value $\eta(I_j)$ for the magnon density within a time window of 50ns at the end of each interval, as illustrated in \cref{fig:mxplotmu5j0}. The resulting magnon density represents one data point in \cref{fig:butterfly_weakStt}b.

The threshold parametric pumping power $h_p^\text{crit}(H_0)$ to excite magnons is occasionally referred to as a \textit{butterfly curve} \cite{patton1979anomalous, guo2014parametric, mohseni2020parametric}. The minimum of the threshold curve lies near the external field corresponding to resonance conditions, $H_\text{FMR}$, which can be approximated by the Kittel formula in \cref{eq:kittel}. Inserting the values of $M_S$ and $\omega_\text{FMR} = \omega_\text{p}/2$ from \cref{tab:simparamstable} results in $H_\text{FMR} \approx \SI{178}{\milli\tesla}$. Note that the Kittel formula is valid for extended thin films, where $L_z/L_{x,y} \rightarrow 0$. In contrast, $L_z/L_{x,y} \approx 0.02$ in the simulated film (see \cref{tab:simparamstable}). This causes a small deviation between the simulated FMR frequency and the Kittel formula. Furthermore, because the lateral lengths of the simulated films are finite, the magnon energy levels are discrete. 
Looking at \cref{fig:butterfly_weakStt}b, we find that the discrete energy levels result in spikes, where it is difficult to excite magnons. The threshold curve can be compared to experimental results by Lauer {\it{et al.}} \cite{lauer2016spin}. 

\subsection{\label{sec:numres_stt}Magnon excitation by STT}

We now proceed to discuss how STT generates magnons. The number of magnons created by STT is a function of the spin accumulation $\mu_S$ and the external field strength $H_0$. The STT effectively controls the damping. The sign of the spin accumulation determines whether the torque acts as a damping-like or anti-damping-like torque. 

\begin{figure}[!htp]
\begin{tikzpicture}
\node (img)  {
\includegraphics[width=1.0\colfigwidth]{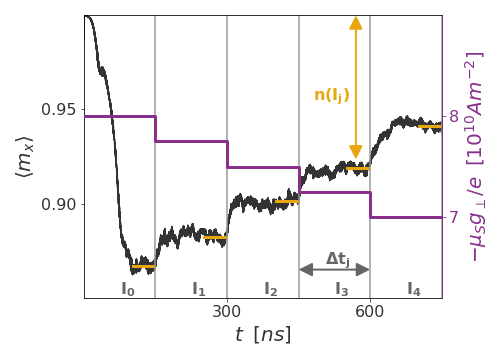}
};
\node[below=of img, node distance=0cm, font=\Large \color{black}, fill=white, 
yshift=2.05cm] {$t [ns]$};
\node[left=of img, node distance=0cm, rotate=90, anchor=center, font=\Large \color{black}, fill=white, 
xshift=2, yshift=-1.4cm] {$\langle m_x \rangle$};
\node[left=of img, node distance=0cm, rotate=90, anchor=center, font=\Large \color{purpleplot}, fill=white, 
xshift=2, yshift=-8.9cm] {$-\mu_S g_\perp / e \; [10^10 Am^{-2}]$};
\end{tikzpicture}
\caption{\label{fig:mxplotsttOfH}
The temporal evolution of the spatial average of the magnetization $\langle m_x \rangle(t)$ during  excitation by STT. 
The simulations start from a magnetization state with random deviations from the uniform state. Magnons are created by the application of a large spin accumulation that results in an anti-damping STT. The spin accumulation decreases in intervals $I_j$ (purple). We compute the temporal average of the magnetization (orange) resulting in the magnon density $\eta(H_0, \mu_S)$. The presented data are an excerpt from the simulations at $\mu_0 H_0 = \SI{100}{\milli\tesla}$. 
}
\end{figure}
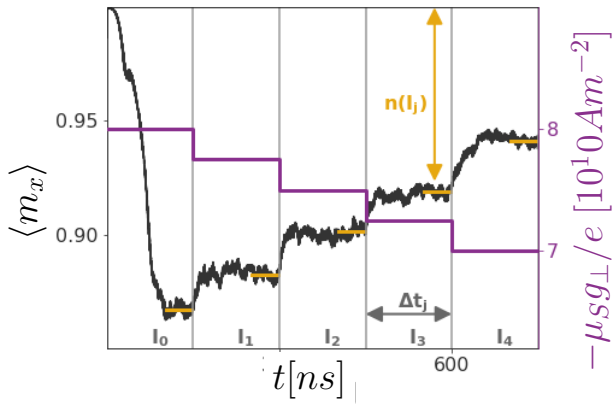

\begin{figure}[h]
\centering
\begin{tikzpicture}
\node (img)  {
\includegraphics[width=1.0\colfigwidth]{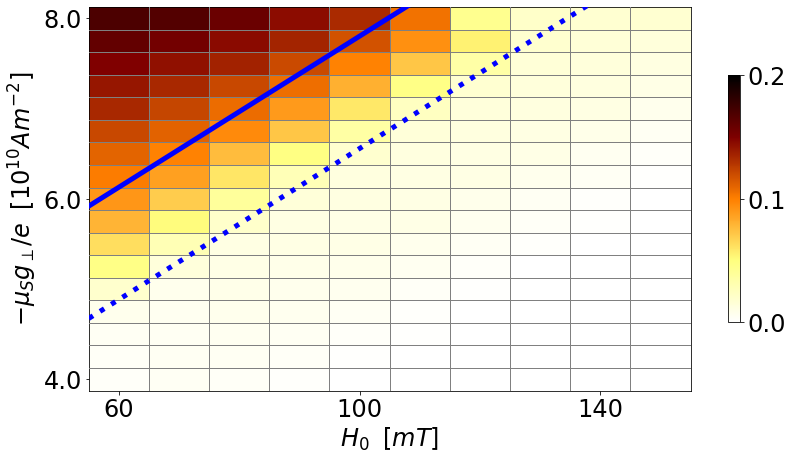}
};
\node[below=of img, node distance=0cm, font=\large \color{black}, fill=white, 
yshift=1.55cm] {$\mu_0 H_0 [mT]$};
\node[left=of img, node distance=0cm, rotate=90, anchor=center, font=\large \color{black}, fill=white, 
xshift=2, yshift=-1.2cm] {$-\mu_S g_\perp / e \; [10^10 Am^{-2}]$};
\end{tikzpicture}
\caption{\label{fig:stt}
The relative magnon density $\eta(H_0, \mu_S)$ during STT excitation of a YIG film as a function of the external magnetic field and spin accumulation. The simulation series for each fixed $H_0$ as in \cref{fig:mxplotsttOfH} represents one column of the figure, where each pixel represents the relative magnon density $\eta(I_j)$. The dashed blue line is the critical spin accumulation $\mu_S^\text{crit}(\vec{k}_\text{STT})$ required to excite magnons, as in \cref{eq:mucrit}. For comparison, the solid blue line shows $\mu_S^\text{crit}(k=0)$.
}
\end{figure}

We use a range of applied bias field strengths to investigate the threshold spin accumulation for exciting magnons by STT. We decrease the strength of the spin accumulation in intervals for each $H_0$-series, as illustrated in \cref{fig:mxplotsttOfH}. The initial magnetization at $t=0$ is set to randomly deviate from the uniform state. An applied spin accumulation ($\mu_S g_{\perp}/e = \SI{-8e10}{\ampere \per \metre \squared}$) causes a strong STT well above the excitation threshold. The torque is present at the entire surface of the film. Next, we gradually decrease the strength of the torque at intervals $I_j$ of duration $\Delta t_j$. The magnetization dynamics is chaotic, and occasionally the system does not easily find a steady state at high current strengths, as shown in \cref{fig:mxplotsttOfH}. The magnon densities $\eta(H_0,\mu_S)$ calculated for the time windows are shown in \cref{fig:stt}. 

We now consider the spin accumulation found by analytical approximations.
The threshold spin accumulation for exciting magnons at wavevector $\vec{k}$ is given in \cref{eq:mucrit}. For the chosen parameters in \cref{tab:simparamstable,tab:yigparamstable}, we find that $\mu_S^\text{crit}(\vec{k})$ reaches its minimum when $\theta_{k_\text{STT}}=0$, $|\vec{k}_\text{STT}| \approx \SI{15.5}{\per \micro \metre}$. 

\subsection{\label{sec:numres_pp_stt}Parallel parametric pumping and weak STT}

The STT is dissipative and effectively changes the Gilbert damping parameter. We therefore expect that the STT changes the threshold of parametric pumping. We investigate this expectation by applying a weak STT ($\mu_S g_{\perp}/e = \pm \SI{1e10}{\ampere \per \metre \squared}$) while performing parametric pumping on the YIG film. The procedure of initialization and interval pumping is similar to that in \cref{sec:numres_pp}. 

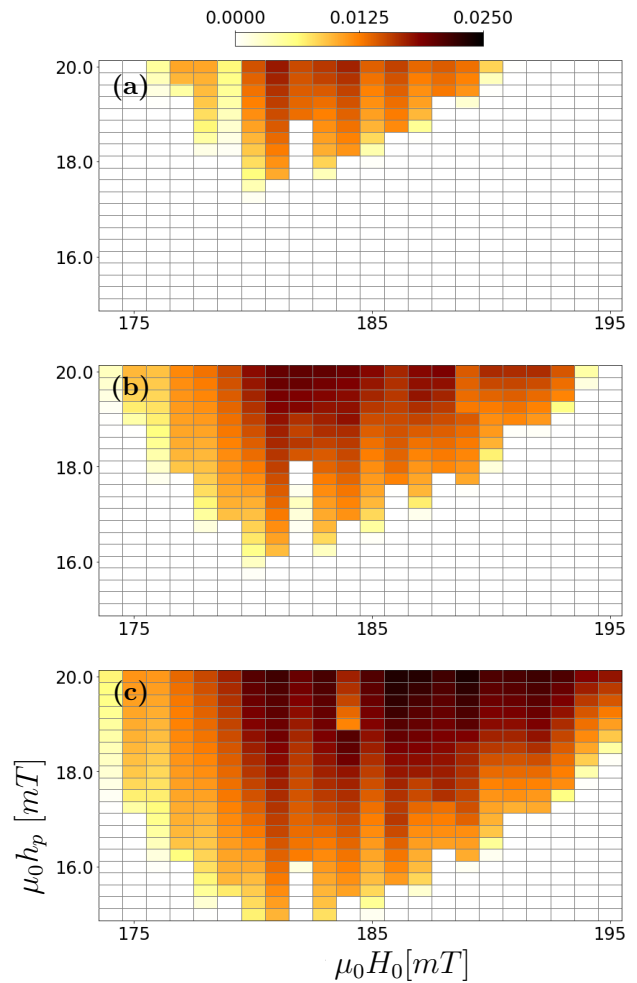
\begin{figure}[!htp]
\centering
\begin{tikzpicture}
\node[above right] (img) at (0,0) {
\includegraphics[width=\colfigwidth]{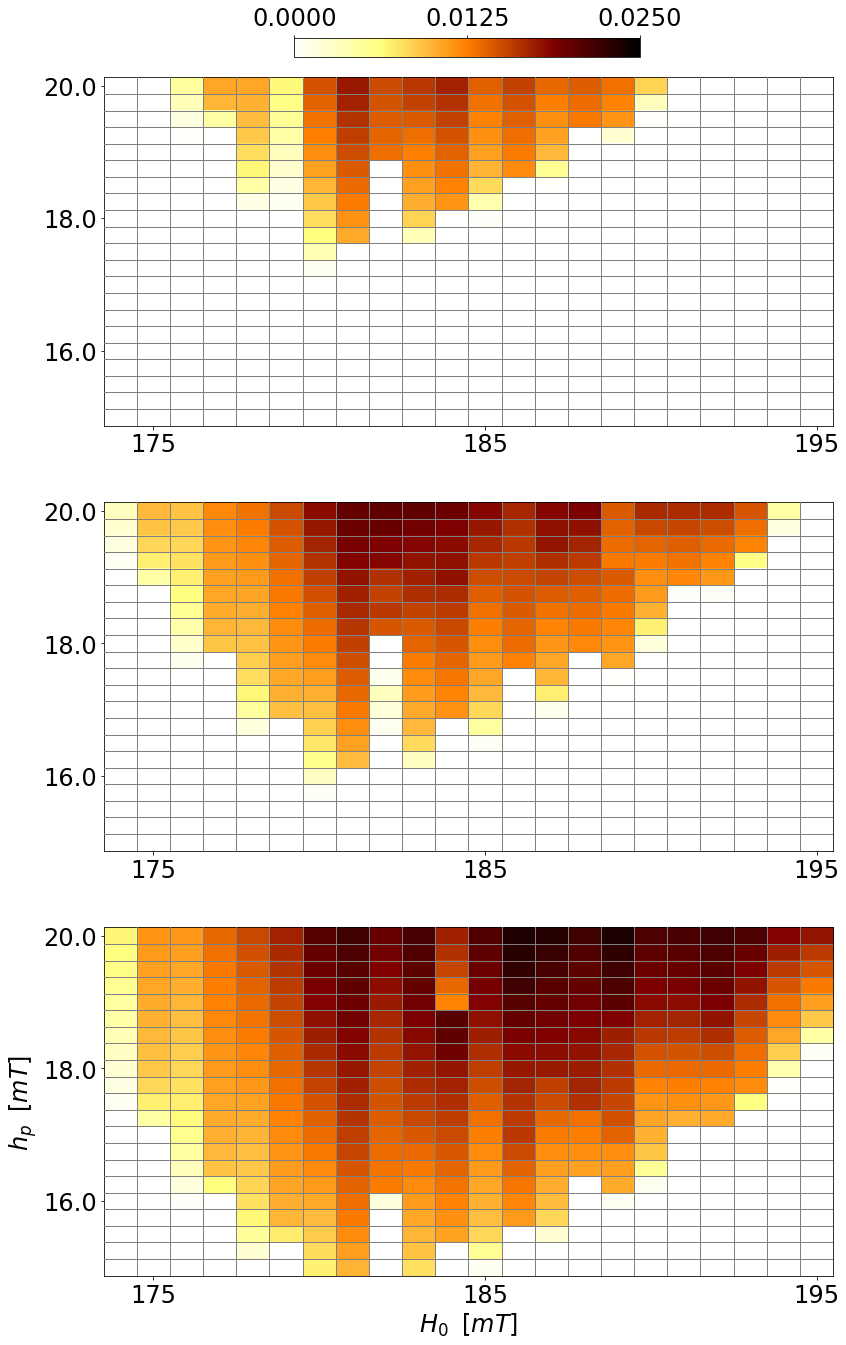}
};
\node at (0.19\figl,1.89\figh) {\textcolor{black}{\normalsize{\bf{(a)}}}};
\node at (0.19\figl,1.25\figh) {\textcolor{black}{\normalsize{\bf{(b)}}}};
\node at (0.19\figl,0.605\figh) {\textcolor{black}{\normalsize{\bf{(c)}}}};
\node[below=of img, node distance=0cm, font=\large \color{black}, fill=white, 
xshift=1cm, yshift=1.5cm] {$\mu_0 H_0 [mT]$};
\node[left=of img, node distance=0cm, rotate=90, anchor=center, font=\large \color{black}, fill=white, 
xshift=-4.5cm, yshift=-1.2cm] {$\mu_0 h_p \; [mT]$};
\end{tikzpicture}
\caption{\label{fig:butterfly_weakStt}
The relative magnon density $\eta(H_0, h_p)$ during the parametric pumping of a thin YIG film. $H_0$ and $h_p$ are the strengths of the external magnetic field and pumping field. 
The simulation series for each fixed $H_0$ (similar to \cref{fig:mxplotmu5j0}) represents one column of the figure, where each pixel represents the relative magnon density $\eta(I_j)$. 
In a) and c), an STT due to the weak spin accumulation $\mu_S$ is applied. \\
a) Damping-like STT, $\mu_S g_{\perp}/e=+\SI{1e10}{\ampere \per \metre \squared}$ \\
b) No STT, $\mu_S=0$ \\
c) Anti-damping-like STT, $\mu_S g_{\perp}/e=-\SI{1e10}{\ampere \per \metre \squared}$
}
\end{figure}

The torque is applied at the entire surface of the film at all times, including during the initialization intervals $I_{-2}$ and $I_{-1}$. 
In our sign convention, a positive (negative) spin accumulation results in a damping-like (anti-damping-like) torque. 
\cref{fig:butterfly_weakStt}a shows the effect of applying a damping-like torque. The magnon instability threshold moves to higher pumping powers since the torque results in a higher effective damping. \cref{fig:butterfly_weakStt}c shows the results of an anti-damping torque, which moves the threshold to lower pumping powers required to excite magnons. Note that in the high magnon density limit, which is the case in the presence of an antidamping-like torque, the magnetization dynamics is highly non-linear. 

\subsection{\label{sec:numres_modes}Excited magnon modes}

\begin{figure}[!htp]
\centering
\subfloat{
\begin{tikzpicture}
\node[above right] (img) at (0,0) {
\includegraphics[width=0.9\colfigwidth]{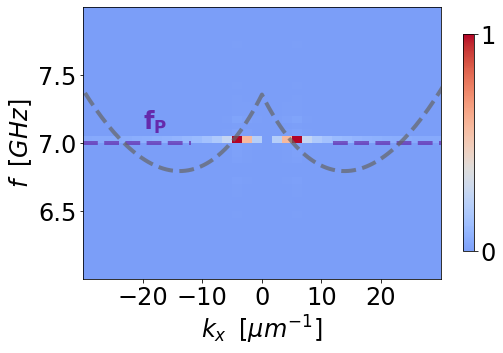}
};
\node at (0.22\figl,0.75\figh) {\textcolor{black}{\normalsize{\bf{(a)}}}};
\end{tikzpicture}
} \\ 
\subfloat{
\begin{tikzpicture}
\node[above right] (img) at (0,0) {
\includegraphics[width=0.9\colfigwidth]{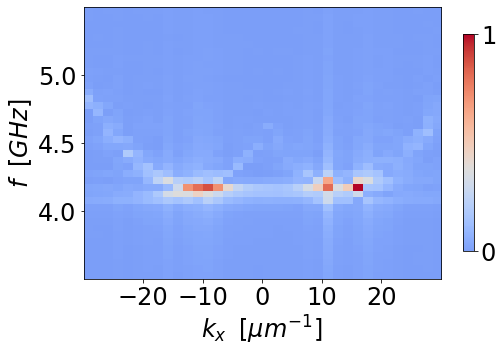}
};
\node at (0.22\figl,0.75\figh) {\textcolor{black}{\normalsize{\bf{(b)}}}};
\end{tikzpicture}
} 
\caption{\label{fig:fourierplots}
The relative distributions of the magnons as a function of frequency and wavevector $k_x$, ($k_y = 0)$. Magnons are excited in the YIG film during 
a) only parametric pumping, $\mu_0 H_0 = \SI{190}{\milli \tesla}$, $\mu_0 h_p = \SI{18}{\milli \tesla}$ and 
b) only STT, $\mu_0 H_0 = \SI{100}{\milli \tesla}$, $\mu_S g_{\perp}/e=-\SI{6.5e10}{\ampere \per \metre \squared}$.
We show the squareroot of the magnon distribution, $\sqrt{\zeta(k_x,k_y)}$ from \cref{eq:magdistribution}, where we disregard edge effects by taking the Fourier transform over the film middle (64 cell edges). The data are normalized with respect to the maximum intensity. In a), the analytical energy dispersion from  \cref{eq:dispersionrel} is illustrated by a gray dashed line, while the pumping frequency is drawn with a purple dashed line. 
}
\end{figure}

Parametric pumping favors the excitation of spin waves with elliptical precession. The dipole interaction dominates the dispersion in the long wavelength limit when the spins precess with an elliptical character. In contrast, the exchange-dominated spin waves with a shorter wavelength have a circular precession. As illustrated in \cref{fig:dispplot}, we expect to mainly pump magnons at relatively long wavelengths, $k=k_{p1}$. To further investigate this aspect, we perform a Fourier transform of the magnetization during parametric pumping and present the results in \cref{fig:fourierplots}a. Our numerical results confirm that the pumped magnons center around $k=k_{p1}$. Keeping the pumping frequency fixed, $k_{p1}$ can be moved to higher (lower) values if we increase (decrease) the strength of the bias field. 

In general, the STT excites spinwaves at a wide range of frequencies and wavevector numbers \cite{demidov2011control}.
For weak currents, the STT is expected to predominantly excite magnons at $k_\text{STT}$, the magnon wavevector that minimizes the spin accumulation given in \cref{eq:mucrit}. Starting from the lowest spin accumulation needed to excite magnons by STT, $k_\text{STT}$ is the wavevector of the first magnons that we expect to excite. $k_\text{STT}$ does not change with the bias field strength, but it depends on the film thickness. From \cref{fig:fourierplots}b, we find that we also excite magnons at neighboring energies. 
We find that the parametric pumping mainly creates dipole-dominated magnons at $k_x = k_{p1} \leq k_\text{min}$. The exchange-dominated magnons at $k_x \geq k_\text{min}$ are more easily excited by using the STT. 

\section{\label{sec:conc}Conclusions}

We have theoretically investigated the density of magnons excited by parametric pumping, STT, and a combination of the two. 
The excitation processes in a thin YIG film are studied by performing micromagnetic simulations.
Spin waves traveling parallel to the bias field direction ("$k_x$-branch") have lower energy than perpendicular spin waves.
During parametric pumping, magnons of elliptic precession are predominantly excited at low wavevector numbers ($k=k_{p1}$) in the $k_x$-branch. 
For STT, we expect the first magnons to be excited at $k=k_\text{STT}$ before being distributed around the energy minimum of the $k_x$-branch. 

We found the critical pumping amplitude or spin accumulation for exciting magnons.
The presence of a damping-like or anti-damping-like STT increases or decreases the threshold power for parametric pumping, depending on the sign of the STT. 
Our computed results are consistent with the measurements in  recent experiments \cite{lauer2016spin}.

\section{Acknowledgments}
The research leading to these results has received funding from the Research Council of Norway through its Centres of Excellence funding scheme, project number 262633, "QuSpin". The simulation work was supported with computational resources from the NTNU Idun infrastructure \cite{sjalander+:2019epic}. A.Q. was supported by the Norwegian Financial Mechanism 2014-2021 under the Polish–Norwegian Research Project NCN GRIEG “2Dtronics” No. 2019/34/H/ST3/00515. 
We thank A. A. Serga and A. Kapelrud for helpful and stimulating discussions.

\newpage

\bibliography{man}{}

\end{document}